\documentclass[aps,twocolumn,floatfix,superscriptaddress]{revtex4-1}

\usepackage{float}
\usepackage{graphics,color}
\usepackage{amsmath}
\usepackage{epsfig}
\usepackage{changebar}
\usepackage{footmisc}
\usepackage{epsfig}
\usepackage{amsfonts}
\usepackage{amssymb}
\usepackage{amsbsy}
\usepackage{soul, xcolor}
\usepackage{braket}
\usepackage{hyperref}
\usepackage[normalem]{ulem}
\setstcolor{blue}

\begin{document}

\title{Emergence of Biaxiality in Nematic Liquid Crystals with Magnetic Inclusions: Some Theoretical Insights}
 
\author{Aditya Vats}
\affiliation{Department of Physics, Indian Institute of Technology Delhi, New Delhi--110016, India.}
\author{Sanjay Puri}
\affiliation{School of Physical Sciences, Jawaharlal Nehru University, New Delhi--110067, India.}
\author{Varsha Banerjee}
\affiliation{Department of Physics, Indian Institute of Technology Delhi, New Delhi--110016, India.}

\begin{abstract}
The biaxial phase in nematic liquid crystals has been elusive for several decades after its prediction in the 1970s.  A recent experimental breakthrough was achieved by Liu et al. [PNAS {\bf 113}, 10479 (2016)] in a liquid crystalline medium with magnetic nanoparticles (MNPs). They exploited the different length-scales of dipolar and magneto-nematic interactions to obtain an equilibrium state where the magnetic moments are at an angle to the nematic director. This tilt introduces a second distinguished direction for orientational ordering or { biaxiality} in the two-component system. Using coarse-grained Ginzburg-Landau free energy models for the nematic and magnetic fields, we provide a theoretical framework which allows for manipulation of morphologies and quantitative estimates of biaxial order.    
\end{abstract}

\maketitle

\section{Introduction}
\label{s1}

Liquid crystal (LC) phases are mesomorphic states between ordinary liquids and crystals. The constituent molecules translate freely as in a liquid while exhibiting long-range orientational order. The simplest LCs are nematic liquid crystals (NLCs), where constituent particles are often rod-like or disc-shaped. The NLC molecules typically orient along a preferred direction ${\bf n}$ called the {\it director}. They exhibit {\it uniaxial} order if the molecular alignment is only about ${\bf n}$. Alternatively, there can be an additional distinguished (secondary) director ${\bf k}$ (perpendicular to ${\bf n}$) for orientational ordering. These are referred to as {\it biaxial} nematic liquid crystals (BNLCs), and were predicted by Freiser in 1970 \cite{Freiser_1970}. BNLCs have been the subject of much experimental and theoretical research \cite{Alben_1973,Alben_1973_2,Straley_1974,Galerne_1998,Bruce_2004,Photinos_2010,Luckhurst_2015}. They are believed to offer significantly improved response times and better viewing characteristics in displays, optical switching and optical imaging as compared to their uniaxial counterparts \cite{Photinos_2010,Luckhurst_2015}. 

The working principle behind LC applications is the {\it Fréedericksz transition}, where the light transmissibility changes when the NLC molecules go from an ordered state to a disordered state \cite{Zannoni_2008,Photinos_2010,Luckhurst_2015,Meyer_2021}. In BNLCs, it was predicted that this transition could occur along more than one direction. However, the experimental detection of thermotropic BNLCs was elusive until 2004, when three groups independently demonstrated the existence of the biaxial phase \cite{Severing_2004,Merkel_2004,Acharya_2004}. It was observed that the Fréedericksz transition about the secondary director is  energetically favorable, yielding light transmission that can potentially be switched on and off more abruptly \cite{Lee_2007,Zannoni_2008, Luckhurst_2015, Photinos_2010,Meyer_2021}. These experiments also revealed that the switching time is at least an order of magnitude faster in BNLCs ($\sim 1$ ms) as compared to uniaxial NLCs ($\sim 15$ ms) \cite{Lee_2007,Luckhurst_2015}. Despite these major advances on the experimental side, the biaxial phase remains a challenge because the ordering of molecules along the secondary director is fragile and easily destroyed by thermal fluctuations \cite{Photinos_2010,Luckhurst_2015}. So the quest for a robust biaxial phase continues. 

A breakthrough in this direction is provided by the recent experiments of Liu et al., where they achieved the elusive biaxial phase by immersing magnetic nanoparticles (MNPs) in an NLC medium \cite{Qliu_PNAS2016}. These fascinating {\it ferronematics} (FNs) were first proposed theoretically in 1970 by Brochard and de Gennes with the purpose of enhancing the magnetic response in NLCs for magneto-optic effects \cite{Broc_1970}. Unfortunately, in experimental samples, MNPs flocculated within tens of minutes due to dipole-dipole interactions \cite{Mert_LCR2017}. It was only four decades later, in 2013, that Mertelj et al. designed the first such stable suspension using barium hexaferrite magnetic nanoplatelets in pentylcyanobiphenyl (5CB) LCs \cite{Mert_Na2013,Mert_LCR2017}. They overcame the challenges of flocculation by cleverly choosing the shape and composition of the MNPs, and a homeotropic MNP-NLC coupling. 

\begin{figure}[H]
\centering                      
\includegraphics[width=0.7\linewidth]{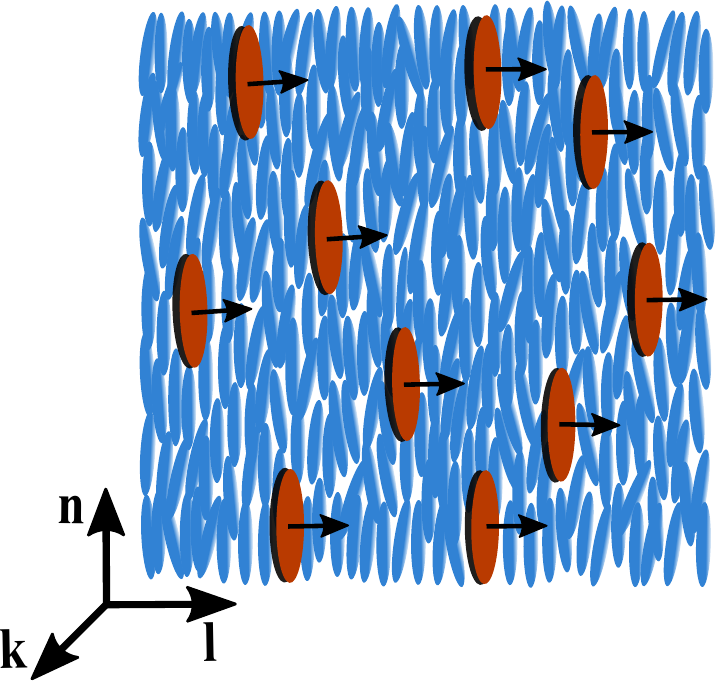} 
\caption{Schematic depicting the orientations of nematic (blue) and magnetic (red) particles for the coupling limit $\gamma<0$.}
\label{f1}
\end{figure}

In their experiments with FNs, Liu et al. \cite{Qliu_PNAS2016} leveraged the different length-scales of dipolar and magneto-nematic interactions to obtain an equilibrium state where the magnetic moment of the MNPs is at an angle to the nematic director ${\bf n}$. Such a coupling introduced an additional direction of order (${\bf k}$) in the perpendicular plane at no additional cost, see the schematic in Fig.~\ref{f1}. Subsequently, the authors confirmed the presence of biaxial order from the absorption spectrum and magnetic hysteresis studies. This development opens up newer horizons for applications of NLCs, and these require theoretical guidance. In this paper, we provide the requisite framework to study biaxial order in FNs. We will demonstrate how the magneto-nematic coupling introduces biaxiality in the system, even though it is absent in the pure NLCs. We also provide quantitative evaluations of biaxiality as a function of the coupling strength, which will be useful for experimentalists.

 This paper is organized as follows. In Sec.~\ref{s2}, we introduce the order parameters and coarse-grained free energy for FNs. In Sec.~\ref{s3}, we present results for the ordering kinetics of FNs, and the development of biaxiality. In Sec.~\ref{s4}, we conclude with a summary and discussion.

\section{Coarse-grained Free Energy for Ferronematics}
\label{s2}

FNs are described in terms of two order parameters: (i) the {\bf Q}-tensor, which contains information about the orientational order of the NLCs, and (ii) the magnetization vector {\bf M}, which gives the average orientation of the magnetic moments of the MNPs. The ${\bf Q}$-tensor is symmetric and traceless, and is given by \cite{Mtram_ArX_2014}:
\begin{equation}
\label{QT}
Q_{ij}=\mathcal{S} n_i n_j + T k_i k_j - (\mathcal{S}+ \tilde{T})\frac{\delta_{ij}}{3} .
\end{equation}
Here, the scalar order parameter $\mathcal{S}$ measures the uniaxial degree of order about the leading eigenvector or the director ${\bf n}$. Further, $\tilde{T}$ is the magnitude of the biaxial order about the secondary director ${\bf k}$. (A system with only uniaxial order has $\tilde{T}=0$. For such a system, the {\it isotropic phase} corresponds to $\mathcal{S}=0$, and the {\it nematic phase} has $\mathcal{S} \ne 0$.) Taking into account the requirements of symmetry and tracelessness, the {\bf Q}-tensor can be expressed in terms of {\it five} independent parameters as follows:
\begin{equation} 
  {\bf Q}=
  \begin{pmatrix}
 -q_1+q_2 & q_3 &  q_4   \\
 q_3 &  -q_1-q_2 & q_5   \\
 q_4 &  q_5 & 2q_1 \\
  \end{pmatrix} .
  \label{Q3}
\end{equation}

To obtain the nematic directors and $\mathcal{S}$, $\tilde{T}$, we choose a frame of reference in which ${\bf Q}$ is diagonal. This provides us the three eigenvalues ($\lambda_3 > \lambda_2 > \lambda_1$), and the corresponding eigenvectors ${\bf n}$, ${\bf k}$, ${\bf l}$. The largest eigenvalue $\lambda_3 = \mathcal{S}$, and the corresponding eigenvector is the primary direction of order ${\bf n}$ \cite{Mtram_ArX_2014, Amit_2010}. We will use a standard measure of biaxial order about the secondary director ${\bf k}$: $\mathcal{T}=(\lambda_2 - \lambda_1)/\lambda_3$ \cite{Photinos_2010, Luckhurst_2015, Amit_2010}, which is proportional to $\tilde{T}$. Naturally,  $\lambda_1 = \lambda_2$ if the system is uniaxial. The degree of biaxiality can also be defined as $\mathcal{B}^2= \{1- 6\text{Tr}({\bf Q}^3)^2/[\text{Tr}({\bf Q}^2)^3]\}$ \cite{Kaiser_1992,kralj_2010}, where $\mathcal{B}^2=0$ for the uniaxial state and $\mathcal{B}^2=1$ for a state with maximum biaxiality. This definition of biaxiality also exploits the difference between two eigenvalues to determine biaxial order, similar to $\mathcal{T}$.

We use the Landau-de Gennes (LdG) approach to write down the phenomenological free energy for this composite system. This is a functional of the order parameter fields {\bf Q}({\bf r}) and {\bf M}({\bf r}) and has three contributions \cite{JP_DG_1995,HPlein_2001,konark_2019,Konark_2019_2,V_2020,V_2021}:
\begin{eqnarray}
\label{LdG_FE}
G[\boldsymbol{Q},\boldsymbol{M}] &=& \int \mbox{d}{\bf r} \left\{ \frac{A}{2}\mbox{Tr}(\boldsymbol{Q}^2)+\frac{C}{3}\mbox{Tr}(\boldsymbol{Q}^3)+\frac{B}{4}[\mbox{Tr}(\boldsymbol{Q}^2)]^2  \right.\nonumber \\ 
&& +\frac{L}{2} \left|\nabla{\bf Q}\right|^2 +\left.\frac{\alpha}{2}\left|\boldsymbol{M}\right|^2+\frac{\beta}{4}\left|\boldsymbol{M}\right|^4 + \frac{\kappa}{2}\left|\nabla \boldsymbol{M}\right|^2 \right.\nonumber \\
&&\left. -\frac{\gamma \mu_0}{2} \sum_{i,j=1}^{3}Q_{ij}M_iM_j\right\}.
\end{eqnarray}
The first four terms in Eq.~(\ref{LdG_FE}) represent the Ginzburg-Landau (GL) free energy for the nematic component with Landau coefficients $A$, $B$, $C$, $L$ having their usual meaning. The next three terms correspond to the GL free energy for the magnetic component. In the GL framework, the gradient terms $|\nabla {\bf Q}|^2$ and $|\nabla {\bf M}|^2$ are essential to capture the effects of elastic interactions \cite{Ravnik_2009,Priestly_2012, Puri_2009,Bray_2002, Hberg_2015}. They penalise local variations in the order parameters -- this surface tension results in the motion of domain boundaries in coarsening kinetics.

The magnitudes of the Landau coefficients determine the scales of order parameter, length and time in the system. For example, $A=A_0(T-T_N)$ and $\alpha=\alpha_0(T-T_M)$ depend on the quench temperature $T$ and the critical temperatures $T_N$, $T_M$. (Here, $A_0$, $\alpha_0$ are material-dependent constants.) A direct estimate of the coefficients can be obtained from experimentally determined quantities like the latent heat, order parameter magnitudes, susceptibilities, etc. \cite{Priestly_2012, Hberg_2015}. However, the current experimental data on FNs is not adequate to provide accurate estimates of these coefficients. The utility of the LdG framework lies primarily in predicting universal behaviors, e.g., power laws and their exponents, scaling variables, etc.

The effect of dopant particles in LCs has been modeled in several previous studies \cite{Li_2006,Lena_2009,Lena_2011,Gorkunov_2011}. These models describe the coupling of the dipole moment of ferroelectric particles with the NLCs at a molecular level. The induced field due to the impurity atoms acts like an aligning field, and enhances orientational order in the NLCs. On a similar footing, the last term in Eq.~(3) is the phenomenological magneto-nematic coupling defined as a dyadic product of ${\bf Q}$ and ${\bf M}$ and the parameter $\gamma$ is the strength of the coupling. It is related to the shape and size of the MNPs and their interaction with the NLCs. This cubic magneto-nematic coupling term \cite{HPlein_2001} enforces the specific orientations of the magnetic and nematic components essential for the emergence of biaxial order in the system \cite{Adam_2021,Qliu_PNAS2016}. A more accurate description of the free energy can be obtained by incorporating dipolar and quadrupolar interactions. This may be required for studies of phase transitions and critical phenomena. As discussed in Ref.~\cite{Konark_2019_2}, these terms may be ignored for dilute ferronematic suspensions.

In their experiments, Liu et al. demonstrated that biaxial order emerges only when ${\bf n}$ and ${\bf M}$ are tilted at an angle. By manipulating the surface functionalization, they could achieve a tilt angle up to $90^{\circ}$. (Their optical absorbance measurements to detect the biaxial phase were carried out for a limited range from 10$^\circ$-65$^{\circ}$.) Motivated by these experiments, we choose $\gamma < 0$ for simplicity, which corresponds to a tilt angle of $90^{\circ}$. In principle, it is possible to modify the coupling term in Eq.~\eqref{LdG_FE} such that ${\bf n}$ and ${\bf M}$ are at an arbitrary angle, but this makes the expression considerably more complicated. The emergence of biaxiality (or the presence of two distinguished directions) in the NLCs for non-zero values of $\gamma<0$ can be understood from the schematic in Fig.~1: Choosing ${\bf M}$ along the positive $x$-axis, the LC molecules can align in two orthogonal directions, say along the $y$-axis and $z$-axis.

In Ref.~\cite{Konark_2019_2}, the authors studied pattern formation in $d=2$ micron-sized ferronematic wells. There, the choice of { $\gamma > 0$} allowed creation of domain walls in the magnetization profile, and stable nematic defects whose location could be manipulated by the magneto-nematic coupling. The present study is a generalization of this framework to $d=3$ to observe the elusive biaxiality.

A few comments regarding the FN free energy are in order: (\romannumeral 1) The state which minimizes the nematic free energy with terms up to order $[\mbox{Tr}(\boldsymbol{Q}^2)]^2$ is always uniaxial. The inclusion of higher-order terms such as $[\mbox{Tr}({\bf Q}^2)]^3$ is necessary for biaxial order in the pure nematic system \cite{Apala_2010,Forest_2000}. (\romannumeral 2) Liu et al. proposed the Frank-free-energy approach to model FNs, which only accounts for the elastic free energy. This simplified framework could not provide a theoretical  understanding of the observed biaxiality. The LdG free energy approach is more generic. It includes the Landau free energy, in addition to the elastic energies. These additional terms are important to identify the state that the LCs would prefer to be in, e.g., uniaxial, biaxial or isotropic \cite{Mtram_ArX_2014,JP_DG_1995}. Further, a quantitative estimate of the biaxial order $\mathcal{T}$ is straightforward from the ${\bf Q}$-tensor.

\section{Ordering Kinetics of Ferronematics}
\label{s3}

\subsection{Time-dependent Ginzburg-Landau Equations}
\label{s3.1}

To obtain the free energy minimum, we study the dissipative dynamics of the FN using the coupled time-dependent Ginzburg-Landau (TDGL) equations:
\begin{equation}
\frac{\partial {\psi}}{\partial t} = -\Gamma_{\psi} \frac{\delta G[\mathbf{Q,M}]}{\delta {\psi}} ,
\label{tdgl}
\end{equation}
where $\psi$ denotes $\mathbf{Q}$ or $\mathbf{M}$. The terms on the right of Eq.~\eqref{tdgl} are the functional derivatives of the free energy functional $G[\mathbf{Q,M}]$ \cite{Hohenberg_1977,Puri_2009, Bray_2002}. This formulation ensures the relaxation of the system to a stable fixed point via the process of domain growth.

A dimensionless form of the TDGL equations can be obtained by introducing the re-scaled variables $\mathbf{Q} = a \mathbf{Q^\prime}$, $\mathbf{M} = b \mathbf{M^\prime}$, $\textbf{r} = \zeta \textbf{r}^\prime$, $t=\eta t^\prime$. The appropriate values of the scale factors are: $a=\sqrt{|A|/2B}$, $b=\sqrt{|\alpha|/\beta}$, $\zeta=\sqrt{\kappa/|\alpha|}$, $\eta=\Gamma_M^{-1}\sqrt{2B/A}$. We drop the primes to obtain the dimensionless evolution equations:
   \begin{eqnarray}
     \frac{1}{\Gamma}\frac{\partial q_1}{\partial t}&=&\xi_1\left[\pm 3q_1- q^23q_1 +{\bar{C}}(6q_1^2-2q_2^2-2q_3^2+q_4^2+q_5^2)\right.\nonumber\\
    && \left.+l\nabla^2 q_1\right] +c_0(-M_1^2-M_2^2 + 2 M_3^2), \label{3d_tdgl_1}\\
     \frac{1}{\Gamma}\frac{\partial q_2}{\partial t}&=&\xi_1\left[\pm q_2-q^2q_2+{\bar{C}}(4q_1q_2+q_4^2-q_5^2)+l\nabla^2 q_2\right]\nonumber \\
     &&+c_0(M_1^2-M_2^2),\label{3d_tdgl_2} \\
     \frac{1}{\Gamma}\frac{\partial q_3}{\partial t}&=&\xi_1\left[\pm q_3-q^2q_3 +{\bar{C}}(-4q_1q_3+2q_4q_5)+l\nabla^2 q_3\right] \nonumber \\ 
     &&+2c_0M_1M_2 \label{3d_tdgl_3},\\
    \frac{1}{\Gamma}\frac{\partial q_4}{\partial t}&=&\xi_1\left[\pm q_4-q^2q_4+{\bar{C}}(2q_1q_4+2q_2q_4+2q_3q_5)\right.\nonumber \\
    &&\left.+l\nabla^2 q_4\right] +2c_0M_1M_3\label{3d_tdgl_4},\\
                \frac{1}{\Gamma}\frac{\partial q_5}{\partial t}&=&\xi_1\left[\pm q_5-q^2q_5 +{\bar{C}}(2q_1q_5-2q_2q_5+2q_3q_4)\right.\nonumber\\
         &&\left.+l\nabla^2 q_5\right] +2c_0M_2M_3,\\\label{3d_tdgl_5}
          \frac{\partial M_1}{\partial t} &=&\xi_2\left[\pm  M_1 -|\mathbf{M}|^2M_1 +\nabla^2 M_1\right]+c_0[(q_2-q_1)M_1\nonumber \\
        &&  +  q_3 M_2 +  q_4 M_3],\label{3d_tdgl_6}\\
                      \frac{\partial M_2}{\partial t} &=&\xi_2\left[\pm M_2 -|\mathbf{M}|^2M_2 +\nabla^2 M_2\right]+c_0[-(q_1+q_2)M_2  \nonumber \\
      &&+  q_3 M_1 +  q_5 M_3],\label{3d_tdgl_7}\\
      \frac{\partial M_3}{\partial t} &=&\xi_2\left[ \pm M_3 - |\mathbf{M}|^2M_3 + \nabla^2 M_3\right]+c_0[2 q_1 M_3  \nonumber\\
      && +  q_4 M_1 + q_5 M_2].\label{3d_tdgl_8} 
   \end{eqnarray}
Here,
\begin{eqnarray}
&&\xi_1 = \dfrac{2 A \beta}{\alpha}\sqrt{\dfrac{A}{2B}},\ \xi_2=\alpha\sqrt{\dfrac{2B}{A}}, \ {\bar{C}}=\dfrac{C}{2\sqrt{2AB}}, \nonumber\\
&&l = \dfrac{L \alpha}{2A\kappa},\ c_0 = \dfrac{\gamma\mu_0}{2}, \ \Gamma=\frac{\alpha\Gamma_Q}{\beta\Gamma_M}\sqrt{\frac{2B}{A}}, \nonumber \\
&& q^2=3q_1^2 + q_2^2 + q_3^2 + q_4^2 + q_5^2 .
\end{eqnarray}

The $\pm$ sign indicates whether the quench temperature is below ($+$) or above ($-$) the critical temperature, say $T_N$ and $T_M$ for the components ${\bf Q}$ and ${\bf M}$, respectively. In this paper, we will study the case with $T < T_N, T_M$. Thus, we consider Eqs.~\eqref{3d_tdgl_1}-\eqref{3d_tdgl_8}, i.e., both ${\bf Q}$ and ${\bf M}$ prefer the ordered state in the absence of coupling ($c_0 = 0$). The parameters $\xi_1$ and $\xi_2$ depend on the magnitudes of ${\bf Q}$ and ${\bf M}$, $l$ is proportional to the relative elastic constant, and $\bar{C}$ determines the order of the transition. The parameter $c_0$ is the magneto-nematic coupling strength, and $\Gamma$ determines the relative time-scales for {\bf Q} and {\bf M } during the evolution process.  Eq.~(13) provides the values of these re-scaled parameters in terms of the Landau coefficients, which depend on the material properties and experimental conditions \cite{Priestly_2012, Hberg_2015}. Notice that different combinations of these coefficients can lead to the same values of the re-scaled parameters. For simplicity, we set $\Gamma = 1$, $\bar{C}=1$, and $l=1$. Unless specified otherwise, the results are presented for $\xi_1=\xi_2=1$.

We have numerically solved Eqs.~(\ref{3d_tdgl_1})-(\ref{3d_tdgl_8}) using the Euler discretization method \cite{kin_Num2009} to determine the evolution of the nematic and magnetic components. The initial fields ${\bf Q}({\bf r},0)$ and ${\bf M}({\bf r},0)$ consisted of small random fluctuations about 0, corresponding to the high-temperature disordered state for both fields. The discretization mesh sizes $\Delta x=1$ and $\Delta t=10^{-4}$ are used in our simulation. Periodic boundary conditions were employed to
\begin{figure}
\centering                      
\includegraphics[width=0.98\linewidth]{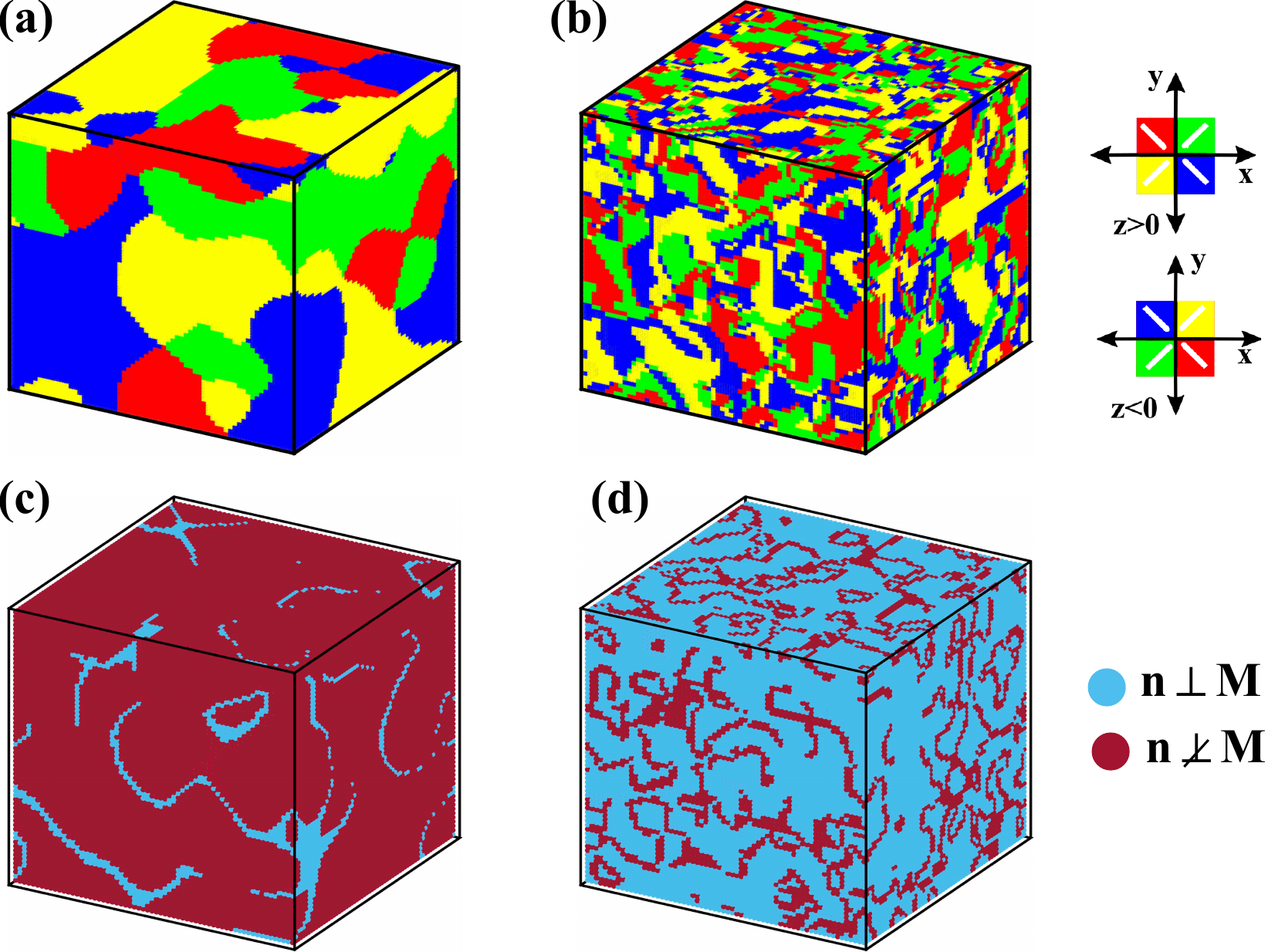} 
\caption{Nematic morphologies for the cases (a) $c_0=0$, and (b) $c_0=-5$ at $t=50$. The regions are colored according to the direction of ${\bf n}$, as shown in the key. The snapshots in (c) $c_0=0$, and (d) $c_0=-5$ depict the regions corresponding to ${\bf n} \perp {\bf M}$ and ${\bf n} \not\perp {\bf M}$.}
\label{f2}
\end{figure}
simulate the bulk behavior and remove edge effects. All statistical results presented here are for the system size $N^3~(N=64)$, averaged over 10 independent runs denoted by $\langle \cdots \rangle$. The evolution of Eqs.~(\ref{3d_tdgl_1})-(\ref{3d_tdgl_8}) provides $\{Q_{ij}\}$ and $M_i$ at all lattice points. The ${\bf Q}$-tensor thus obtained is symmetric and traceless, but not necessarily diagonal. The physically relevant quantities ${\bf n}, {\bf k}, \mathcal{S}$ and $\mathcal{T}$ can be obtained from ${\bf Q}$, refer text following Eq.~(\ref{QT}).
              
Starting with identical random initial conditions, Fig.~\ref{f2} shows evolution snapshots of the nematic morphology (${\bf n}$) at $t=50$ for (a) $c_0=0$ and (b) $c_0=-5$. The ${\bf n}$-field has inversion symmetry, so the orientation at each point on the cubic grid can be represented by one of the 4 colors shown in the key. The growth of domains is faster in the uncoupled system as compared to the FN. Recall that the magneto-nematic coupling parameter $\gamma<0$ coerces ${\bf n}$ to be perpendicular to ${\bf M}$. The lower panel again shows the ${\bf n}$-field at $t=50$ for (c) $c_0=0$, and (d) $c_0=-5$. In these sub-figures, regions with ${\bf n} \perp {\bf M}$ are identified as those where the dot product $|{\bf n}\cdot {\bf M}| < 0.05$. In (c), both ${\bf n}$ and ${\bf M}$ undergo ordering but their relative directions are not constrained. On the other hand, in (d) the magneto-nematic coupling enforces ${\bf n}\perp{\bf M}$.

\subsection{Emergence of Biaxiality}
\label{s3.2}

\begin{figure}
\centering                      
\includegraphics[width=0.98\linewidth]{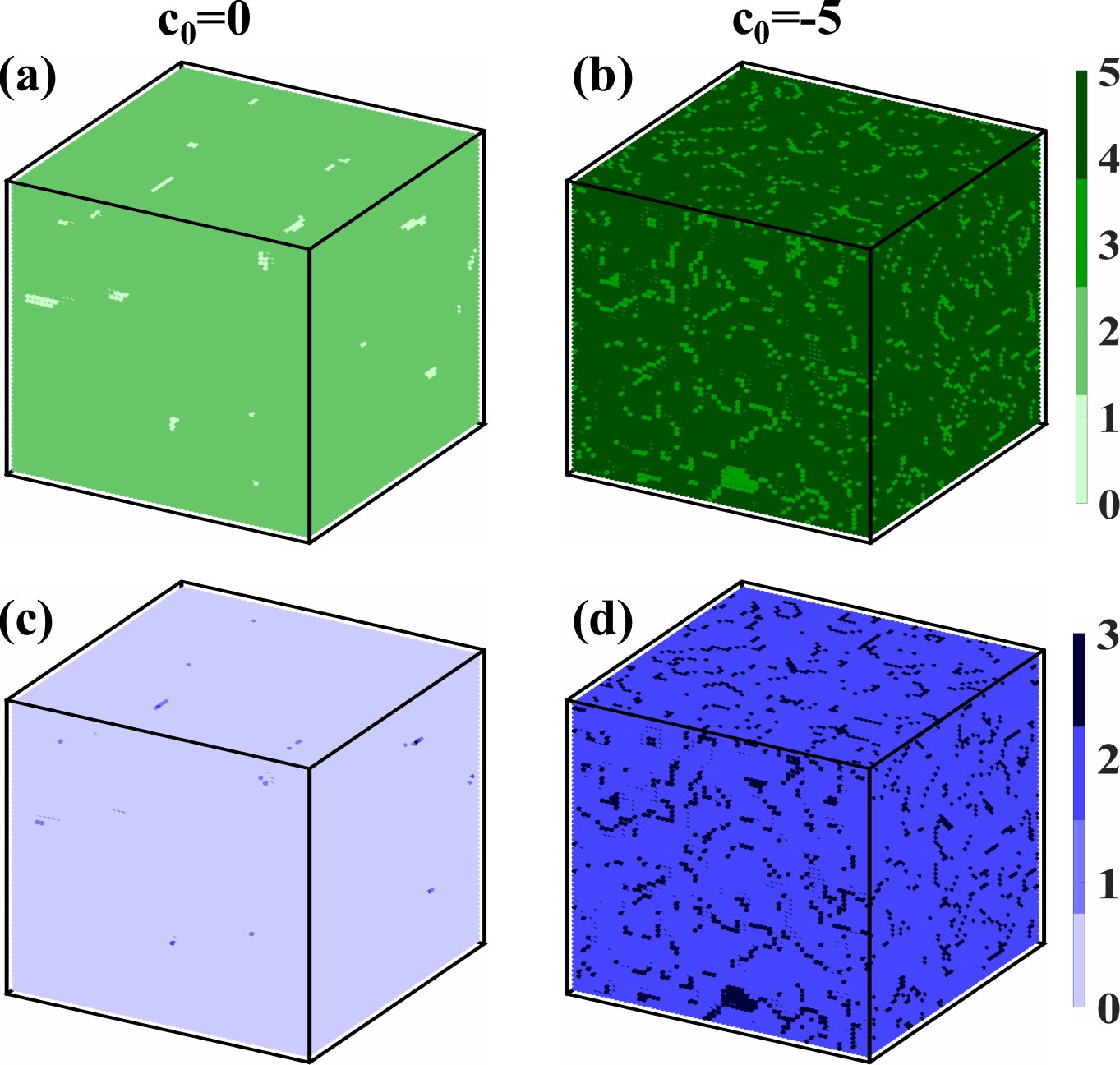} 
\caption{Morphologies of the $\mathcal{S}$-field for (a) $c_0=0$, and (b) $c_0=-5.0$ at $t=50$. The regions are colored according to the magnitude of $\mathcal{S}$. The corresponding $\mathcal{T}$-field is shown below in (c) $c_0=0$, and (d) $c_0=-5.0$.}
\label{f3}
\end{figure}

Let us now demonstrate that the ${\bf Q}$-field in Fig.~\ref{f2} becomes biaxial when the coupling is introduced. Uniaxial LCs have average orientational order along the (primary) director ${\bf n}$.  Additional orientational order in the perpendicular plane signifies the presence of yet another (secondary) director ${\bf k}$ leading to biaxiality in the system \cite{Mtram_ArX_2014}. In Figs.~\ref{f3}(a)-(b), we plot the order parameter $\mathcal{S}$ of the ${\bf n}$-field at $t=50$ for $c_0 = 0, -5$. The darker regions in the snapshots denote regions with higher values of $\mathcal{S}$. Clearly, the ${\bf n}$-field is significantly ordered in both cases. In Figs.~\ref{f3}(c)-(d), we plot the corresponding order parameter $\mathcal{T}$ of the ${\bf k}$-field (secondary director). In this case, we see that there is significant order only when the magneto-nematic coupling is turned on.

\begin{figure}[H]
\centering                      
\includegraphics[width=\linewidth]{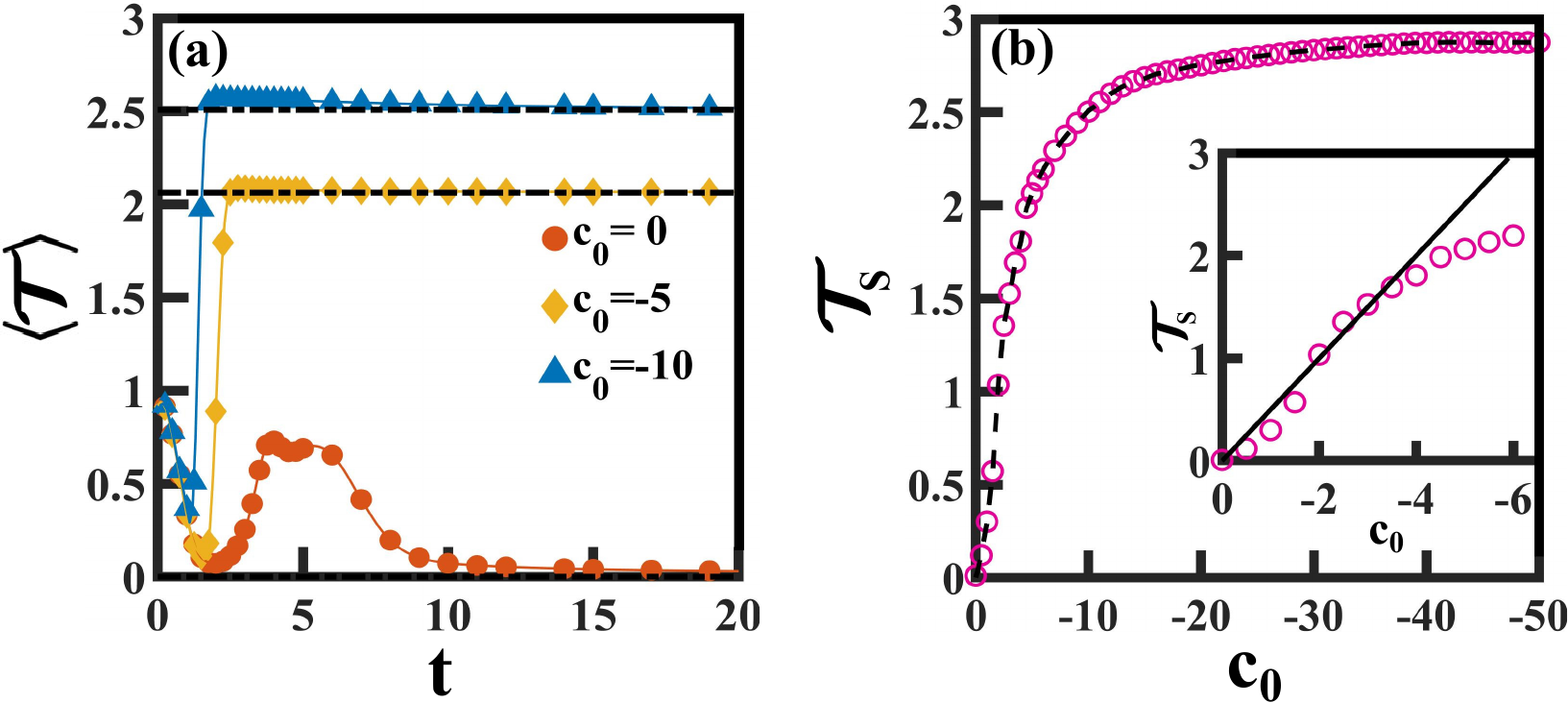} 
\caption{(a) Plot of the average biaxiality parameter, $\langle \mathcal{T} \rangle$ vs. $t$, for different values of $c_0$. The dashed lines correspond to the fixed-point values of $\mathcal{T}$ for $c_0=-5,-10$. (b) Plot of saturation value of biaxiality parameter $\mathcal{T}_s$ vs. $c_0$. The dashed line denotes the fixed-point values of $\mathcal{T}$, obtained numerically from the TDGL equations. The inset shows the behavior for small $c_0$. The solid line denotes the result in Eq.~\eqref{pert2} with $\bar{C}=1$.}
\label{f4}
\end{figure}

Next, we estimate the average biaxiality parameter $\langle \mathcal{T} \rangle$. This is obtained by spatially averaging $\mathcal{T} ({\bf r},t)$ for each run, and then averaging over independent runs.  Fig.~\ref{f4}(a) shows $\langle\mathcal{T} \rangle$ vs. $t$ for different values of $c_0$. For the uncoupled limit $c_0=0$, $\langle \mathcal{T} \rangle \simeq 0$ after the initial transients, signifying relaxation to the uniaxial state.
For $c_0 < 0$, $\langle\mathcal{T}\rangle$ grows and saturates to $\mathcal{T}_s$ at late times. (We have checked this for values starting from $c_0=-0.05$.) The saturation values are obtained from the fixed point solutions ${\bf Q^*}$ and ${\bf M^*}$ of the TDGL equations. These can be obtained by first setting $\partial /\partial t = 0$ and $\nabla^2 = 0$ in Eqs.~(\ref{3d_tdgl_1})-(\ref{3d_tdgl_8}), and solving the coupled equations numerically via the Newton-Raphson method \cite{kin_Num2009}. The relevant equations are:
 \begin{eqnarray}
&& \xi_1\left[\pm 3q_1- q^23q_1 +{\bar{C}}(6q_1^2-2q_2^2-2q_3^2+q_4^2+q_5^2)\right] \nonumber \\
&& +c_0(-M_1^2-M_2^2 + 2 M_3^2)=0, \label{FF_1}\\
&& \xi_1\left[\pm q_2-q^2q_2+{\bar{C}}(4q_1q_2+q_4^2-q_5^2)\right]\nonumber \\
&& +c_0(M_1^2-M_2^2)=0,\label{FF_2} \\
&& \xi_1\left[\pm q_3-q^2q_3 +{\bar{C}}(-4q_1q_3+2q_4q_5)\right] \nonumber \\ 
&& +2c_0M_1M_2=0\label{FF_3},\\
&& \xi_1 \left[\pm q_4-q^2q_4+{\bar{C}}(2q_1q_4+2q_2q_4+2q_3q_5) \right] \nonumber \\
&& +2c_0M_1M_3=0\label{FF_4},\\
&& \xi_1\left[\pm q_5-q^2q_5 +{\bar{C}}(2q_1q_5-2q_2q_5+2q_3q_4) \right] \nonumber\\
&& +2c_0M_2M_3=0,\\\label{FF_5}
&& \xi_2\left[\pm  M_1 -|\mathbf{M}|^2M_1 \right]+c_0[(q_2-q_1)M_1\nonumber \\
&& +  q_3 M_2 +  q_4 M_3]=0,\label{FF_6}\\
&& \xi_2\left[\pm M_2 -|\mathbf{M}|^2M_2 \right]+c_0[-(q_1+q_2)M_2  \nonumber \\
&& + q_3 M_1 +  q_5 M_3]=0,\label{FF_7}\\
&& \xi_2\left[ \pm M_3 - |\mathbf{M}|^2M_3 \right]+c_0[2 q_1 M_3  \nonumber\\
&& +  q_4 M_1 + q_5 M_2]=0.\label{FF_8} 
\end{eqnarray} 
The dashed horizontal lines in Fig.~\ref{f4}(a) denote the fixed-point values obtained numerically from Eqs.~(\ref{FF_1})-(\ref{FF_8}). Next, we obtain the relation between $\mathcal{T}_s$ and the magneto-nematic coupling strength. Fig.~\ref{f4}(b) shows the variation of $\mathcal{T}_s$ vs. $c_0$. Notice that $\mathcal{T}_s$ increases for small $c_0$ and then saturates for larger values of $c_0$.

The small-$c_0$ dependence of $\mathcal{T}$ for $c_0<0$ can be obtained analytically using a perturbative approach as follows. Let ${\bf Q^*}={\bf Q^*_0} + \Delta{\bf Q}$ and ${\bf M^*}={\bf M^*_0} + \Delta{\bf M}$, where (${\bf Q^*_0},{\bf M^*_0}$) are the fixed points of the uncoupled equations ($c_0=0$). Without loss of generality, we use rotational invariance to make the choice
\begin{eqnarray}
{\bf Q^*_0}=
\begin{pmatrix}
-q_1^* & 0 &  0   \\
0 &  -q_1^* & 0   \\
0 &  0 & 2q_1^* \\
\end{pmatrix}, \quad {\bf M^*_0}=\left(1,0,0\right) ,
\label{QM0}
\end{eqnarray} 
where
\begin{eqnarray}
q_1^*= \frac{2\bar{C}+\sqrt{4{\bar{C}}^2+12}}{6} .
\end{eqnarray}
This corresponds to ${\bf n^*_0}$ pointing along the $z$-axis, and ${\bf M^*_0}$ pointing along the $+x$-axis, { i.e., ${\bf n^*_0} \perp {\bf M^*_0}$. Thus, the base state for our expansion is only valid for $c_0 < 0$. For $c_0 > 0$, a suitable base state would have ${\bf n^*_0} \parallel {\bf M^*_0}$.} The expressions for $(\Delta{\bf Q},\Delta{\bf M})$, correct to $O(c_0)$, can be obtained from Eqs.~(\ref{FF_1})-(\ref{FF_8}) with $\xi_1=\xi_2=1$:
\begin{eqnarray}
{\bf \Delta Q}&=&
\begin{pmatrix}
-\dfrac{(3+2\bar{C})c_0}{6\bar{C}q_1^*(1+\bar{C}q_1^*)}& 0 &  0   \\
0 & \dfrac{(4\bar{C}q_1^*+3)c_0}{6\bar{C}q_1^*(1+\bar{C}q_1^*)}  & 0   \\
0 &  0 & -\dfrac{c_0}{3(1+\bar{C}q_1^*)} \\
\end{pmatrix} , \nonumber \\
{\Delta {\bf M}}&=&\left ( -\dfrac{c_0q_1^*}{2},0,0\right).
\label{DQM}
\end{eqnarray}

From the ${\bf Q}$-tensor, the small-$c_0$ dependence of $\mathcal{S}$ and $\mathcal{T}$ can be obtained as
\begin{eqnarray}
\mathcal{S}&=&\dfrac{(6+4\bar{C}^2)q_1^*+2\bar{C}-c_0}{3(1+\bar{C}q_1^*)}+\mathcal{O}(c_0^2) ,\label{pert1} \\
\mathcal{T}&=&-\frac{3(1+\bar{C}q_1^*)c_0}{\bar{C}q_1^*(6q_1^*+4\bar{C}^2q_1^*+2\bar{C})}+\mathcal{O}(c_0^2) .
\label{pert2}
\end{eqnarray}
( We stress that Eqs.~(\ref{pert1})-(\ref{pert2}) are only valid for $c_0 < 0$ due to our choice of the unperturbed state. For $c_0 > 0$, an exact numerical solution of Eqs.~(\ref{FF_1})-(\ref{FF_8}), where we carefully consider all possible roots, shows that $\mathcal {T} = 0$.) The solid line in the inset of Fig.~\ref{f4}(b) denotes $\mathcal{T}$ vs. $c_0$ from Eq.~\eqref{pert2} with $\bar{C}=1$. There is very good agreement with the numerical results up to $c_0 \simeq -4.0$.
\begin{figure}[H]
        \centering
        \includegraphics[width=0.72\linewidth]{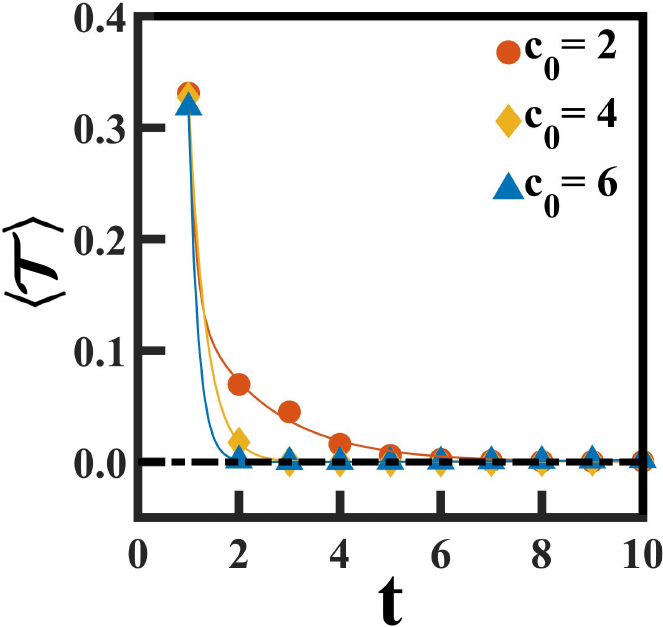} 
                \caption{Plot of average biaxiality parameter, $\langle\mathcal{T} \rangle$ vs. $t$, for different values of positive $c_0$.}
        \label{f5}
\end{figure}
We also demonstrate that the equilibrium morphologies for $c_0 > 0$ are uniaxial in nature. In Fig.~\ref{f5}, we show the time-dependence of the average biaxiality parameter $\langle \mathcal{T} \rangle$ for $c_0 = 2.0, 4.0, 6.0$, evaluated from Eqs.~\eqref{3d_tdgl_1}-\eqref{3d_tdgl_8}. The dashed line denotes the fixed-point value of $\mathcal{T} = 0$, obtained by a Newton-Raphson solution of Eqs.~(\ref{FF_1})-(\ref{FF_8}). Clearly, $\langle \mathcal{T} \rangle \rightarrow 0$ at late times, confirming uniaxial order for $c_0 > 0$.

\section{Summary and Discussion}
\label{s4}

To conclude, we have presented a framework that explains the emergence of biaxiality due to the magneto-nematic coupling in nematic liquid crystals with magnetic inclusions or {\it ferronematics}. This topic has generated interest because of its potential application in the multi-billion dollar LC display industry. Further excitement has resulted after the benchmarking experiments of Liu et al. \cite{Qliu_PNAS2016}, which demonstrated the emergence of the elusive biaxial order in FNs. Our framework to guide experiments in these unique systems with the twin properties of magnetism and biaxiality is therefore very timely. We have used coarse-grained Landau-de Gennes free energies and a time-dependent Ginzburg-Landau formulation to explore the free energy minima of this coupled system. The different feature is the inclusion of a coupling parameter $c_0<0$ due to which the FN relaxes to a state where ${\bf n}\perp {\bf M}$. This choice is crucial for the emergence of biaxiality in our study, and is also consistent with the experiments of Liu et al. \cite{Qliu_PNAS2016}. Our formulation provides a quantitative evaluation of biaxiality and its dependence on the magneto-nematic coupling strength. The latter, in principle, can be manipulated in the laboratory. We hope that this quantification will enable more systematic experiments. 

In a related context, we also mention the earlier experiments of Mertelj et al. which created the first stable FN with enhanced magnetic response \cite{Mert_Na2013,V_2020,V_2021}. In the Mertelj experiments, the equilibrium state of the FN had ${\bf n} \parallel {\bf M}$. The work of Mertelj et al. formed the basis of the experiments by Liu et al. Our theoretical formulation with $c_0 > 0$ mimics the key results of the experiments of Mertelj et al. This choice promotes alignment of the nematic and magnetic order parameters \cite{V_2020,V_2021}. However, we emphasize that this class of systems shows uniaxial behavior. Therefore, by manipulating model parameters, our formulation allows for tailoring morphologies as well as biaxiality. FNs are of great fundamental and technological interest, and much remains to be understood regarding their equilibrium and non-equilibrium properties. Our study is a modest step in this direction. We hope that it will provoke joint experimental and theoretical investigations in this area.

\subsection*{Acknowledgments}

AV acknowledges UGC, India for a research fellowship. VB acknowledges DST-UKIERI and DST India for research grants. AV and VB gratefully acknowledge the HPC facility of IIT Delhi for computational resources. We are grateful to the referees for their useful comments.

\bibliography{embiax}

\begin{thebibliography}{42}%
\makeatletter
\providecommand \@ifxundefined [1]{%
 \@ifx{#1\undefined}
}%
\providecommand \@ifnum [1]{%
 \ifnum #1\expandafter \@firstoftwo
 \else \expandafter \@secondoftwo
 \fi
}%
\providecommand \@ifx [1]{%
 \ifx #1\expandafter \@firstoftwo
 \else \expandafter \@secondoftwo
 \fi
}%
\providecommand \natexlab [1]{#1}%
\providecommand \enquote  [1]{``#1''}%
\providecommand \bibnamefont  [1]{#1}%
\providecommand \bibfnamefont [1]{#1}%
\providecommand \citenamefont [1]{#1}%
\providecommand \href@noop [0]{\@secondoftwo}%
\providecommand \href [0]{\begingroup \@sanitize@url \@href}%
\providecommand \@href[1]{\@@startlink{#1}\@@href}%
\providecommand \@@href[1]{\endgroup#1\@@endlink}%
\providecommand \@sanitize@url [0]{\catcode `\\12\catcode `\$12\catcode
  `\&12\catcode `\#12\catcode `\^12\catcode `\_12\catcode `\%12\relax}%
\providecommand \@@startlink[1]{}%
\providecommand \@@endlink[0]{}%
\providecommand \url  [0]{\begingroup\@sanitize@url \@url }%
\providecommand \@url [1]{\endgroup\@href {#1}{\urlprefix }}%
\providecommand \urlprefix  [0]{URL }%
\providecommand \Eprint [0]{\href }%
\providecommand \doibase [0]{http://dx.doi.org/}%
\providecommand \selectlanguage [0]{\@gobble}%
\providecommand \bibinfo  [0]{\@secondoftwo}%
\providecommand \bibfield  [0]{\@secondoftwo}%
\providecommand \translation [1]{[#1]}%
\providecommand \BibitemOpen [0]{}%
\providecommand \bibitemStop [0]{}%
\providecommand \bibitemNoStop [0]{.\EOS\space}%
\providecommand \EOS [0]{\spacefactor3000\relax}%
\providecommand \BibitemShut  [1]{\csname bibitem#1\endcsname}%
\let\auto@bib@innerbib\@empty
\bibitem [{\citenamefont {Freiser}(1970)}]{Freiser_1970}%
  \BibitemOpen
  \bibfield  {author} {\bibinfo {author} {\bibfnamefont {M.~J.}\ \bibnamefont
  {Freiser}},\ }\href@noop {} {\bibfield  {journal} {\bibinfo  {journal} {Phys.
  Rev. Lett.}\ }\textbf {\bibinfo {volume} {24}},\ \bibinfo {pages} {1041}
  (\bibinfo {year} {1970})}\BibitemShut {NoStop}%
\bibitem [{\citenamefont {Alben}(1973{\natexlab{a}})}]{Alben_1973}%
  \BibitemOpen
  \bibfield  {author} {\bibinfo {author} {\bibfnamefont {R.}~\bibnamefont
  {Alben}},\ }\href@noop {} {\bibfield  {journal} {\bibinfo  {journal} {Phys.
  Rev. Lett.}\ }\textbf {\bibinfo {volume} {30}},\ \bibinfo {pages} {778}
  (\bibinfo {year} {1973}{\natexlab{a}})}\BibitemShut {NoStop}%
\bibitem [{\citenamefont {Alben}(1973{\natexlab{b}})}]{Alben_1973_2}%
  \BibitemOpen
  \bibfield  {author} {\bibinfo {author} {\bibfnamefont {R.}~\bibnamefont
  {Alben}},\ }\href@noop {} {\bibfield  {journal} {\bibinfo  {journal} {J.
  Chem. Phys.}\ }\textbf {\bibinfo {volume} {59}},\ \bibinfo {pages} {4299}
  (\bibinfo {year} {1973}{\natexlab{b}})}\BibitemShut {NoStop}%
\bibitem [{\citenamefont {Straley}(1974)}]{Straley_1974}%
  \BibitemOpen
  \bibfield  {author} {\bibinfo {author} {\bibfnamefont {J.~P.}\ \bibnamefont
  {Straley}},\ }\href@noop {} {\bibfield  {journal} {\bibinfo  {journal} {Phys.
  Rev. A}\ }\textbf {\bibinfo {volume} {10}},\ \bibinfo {pages} {1881}
  (\bibinfo {year} {1974})}\BibitemShut {NoStop}%
\bibitem [{\citenamefont {Galerne}(1998)}]{Galerne_1998}%
  \BibitemOpen
  \bibfield  {author} {\bibinfo {author} {\bibfnamefont {Y.}~\bibnamefont
  {Galerne}},\ }\href@noop {} {\bibfield  {journal} {\bibinfo  {journal} {Mol.
  Cryst. Liq. Cryst.}\ }\textbf {\bibinfo {volume} {323}},\ \bibinfo {pages}
  {211} (\bibinfo {year} {1998})}\BibitemShut {NoStop}%
\bibitem [{\citenamefont {Bruce}(2004)}]{Bruce_2004}%
  \BibitemOpen
  \bibfield  {author} {\bibinfo {author} {\bibfnamefont {D.~W.}\ \bibnamefont
  {Bruce}},\ }\href@noop {} {\bibfield  {journal} {\bibinfo  {journal} {Chem.
  Rec.}\ }\textbf {\bibinfo {volume} {4}},\ \bibinfo {pages} {10} (\bibinfo
  {year} {2004})}\BibitemShut {NoStop}%
\bibitem [{\citenamefont {Tschierske}\ and\ \citenamefont
  {Photinos}(2010)}]{Photinos_2010}%
  \BibitemOpen
  \bibfield  {author} {\bibinfo {author} {\bibfnamefont {C.}~\bibnamefont
  {Tschierske}}\ and\ \bibinfo {author} {\bibfnamefont {D.~J.}\ \bibnamefont
  {Photinos}},\ }\href@noop {} {\bibfield  {journal} {\bibinfo  {journal} {J.
  Mater. Chem.}\ }\textbf {\bibinfo {volume} {20}},\ \bibinfo {pages} {4263}
  (\bibinfo {year} {2010})}\BibitemShut {NoStop}%
\bibitem [{\citenamefont {Luckhurst}\ and\ \citenamefont
  {Sluckin}(2015)}]{Luckhurst_2015}%
  \BibitemOpen
  \bibfield  {author} {\bibinfo {author} {\bibfnamefont {G.~R.}\ \bibnamefont
  {Luckhurst}}\ and\ \bibinfo {author} {\bibfnamefont {T.~J.}\ \bibnamefont
  {Sluckin}},\ }\href@noop {} {\emph {\bibinfo {title} {Biaxial nematic liquid
  crystals: theory, simulation and experiment}}}\ (\bibinfo  {publisher} {John
  Wiley \& Sons},\ \bibinfo {year} {2015})\BibitemShut {NoStop}%
\bibitem [{\citenamefont {Berardi}\ \emph {et~al.}(2008)\citenamefont
  {Berardi}, \citenamefont {Muccioli},\ and\ \citenamefont
  {Zannoni}}]{Zannoni_2008}%
  \BibitemOpen
  \bibfield  {author} {\bibinfo {author} {\bibfnamefont {R.}~\bibnamefont
  {Berardi}}, \bibinfo {author} {\bibfnamefont {L.}~\bibnamefont {Muccioli}}, \
  and\ \bibinfo {author} {\bibfnamefont {C.}~\bibnamefont {Zannoni}},\
  }\href@noop {} {\bibfield  {journal} {\bibinfo  {journal} {J. Chem. Phys.}\
  }\textbf {\bibinfo {volume} {128}},\ \bibinfo {pages} {024905} (\bibinfo
  {year} {2008})}\BibitemShut {NoStop}%
\bibitem [{\citenamefont {Meyer}\ \emph {et~al.}(2021)\citenamefont {Meyer},
  \citenamefont {Davidson}, \citenamefont {Constantin}, \citenamefont {Sergan}
  \emph {et~al.}}]{Meyer_2021}%
  \BibitemOpen
  \bibfield  {author} {\bibinfo {author} {\bibfnamefont {C.}~\bibnamefont
  {Meyer}}, \bibinfo {author} {\bibfnamefont {P.}~\bibnamefont {Davidson}},
  \bibinfo {author} {\bibfnamefont {D.}~\bibnamefont {Constantin}}, \bibinfo
  {author} {\bibfnamefont {V.}~\bibnamefont {Sergan}},  \emph {et~al.},\
  }\href@noop {} {\bibfield  {journal} {\bibinfo  {journal} {Phys. Rev. X}\
  }\textbf {\bibinfo {volume} {11}},\ \bibinfo {pages} {031012} (\bibinfo
  {year} {2021})}\BibitemShut {NoStop}%
\bibitem [{\citenamefont {Severing}\ and\ \citenamefont
  {Saalw{\"a}chter}(2004)}]{Severing_2004}%
  \BibitemOpen
  \bibfield  {author} {\bibinfo {author} {\bibfnamefont {K.}~\bibnamefont
  {Severing}}\ and\ \bibinfo {author} {\bibfnamefont {K.}~\bibnamefont
  {Saalw{\"a}chter}},\ }\href@noop {} {\bibfield  {journal} {\bibinfo
  {journal} {Phys.l Rev. Lett.}\ }\textbf {\bibinfo {volume} {92}},\ \bibinfo
  {pages} {125501} (\bibinfo {year} {2004})}\BibitemShut {NoStop}%
\bibitem [{\citenamefont {Merkel}\ \emph {et~al.}(2004)\citenamefont {Merkel},
  \citenamefont {Kocot}, \citenamefont {Vij}, \citenamefont {Korlacki},
  \citenamefont {Mehl},\ and\ \citenamefont {Meyer}}]{Merkel_2004}%
  \BibitemOpen
  \bibfield  {author} {\bibinfo {author} {\bibfnamefont {K.}~\bibnamefont
  {Merkel}}, \bibinfo {author} {\bibfnamefont {A.}~\bibnamefont {Kocot}},
  \bibinfo {author} {\bibfnamefont {J.}~\bibnamefont {Vij}}, \bibinfo {author}
  {\bibfnamefont {R.}~\bibnamefont {Korlacki}}, \bibinfo {author}
  {\bibfnamefont {G.}~\bibnamefont {Mehl}}, \ and\ \bibinfo {author}
  {\bibfnamefont {T.}~\bibnamefont {Meyer}},\ }\href@noop {} {\bibfield
  {journal} {\bibinfo  {journal} {Phys. Rev. Lett.}\ }\textbf {\bibinfo
  {volume} {93}},\ \bibinfo {pages} {237801} (\bibinfo {year}
  {2004})}\BibitemShut {NoStop}%
\bibitem [{\citenamefont {Acharya}\ \emph {et~al.}(2004)\citenamefont
  {Acharya}, \citenamefont {Primak},\ and\ \citenamefont
  {Kumar}}]{Acharya_2004}%
  \BibitemOpen
  \bibfield  {author} {\bibinfo {author} {\bibfnamefont {B.~R.}\ \bibnamefont
  {Acharya}}, \bibinfo {author} {\bibfnamefont {A.}~\bibnamefont {Primak}}, \
  and\ \bibinfo {author} {\bibfnamefont {S.}~\bibnamefont {Kumar}},\
  }\href@noop {} {\bibfield  {journal} {\bibinfo  {journal} {Phys. Rev. Lett.}\
  }\textbf {\bibinfo {volume} {92}},\ \bibinfo {pages} {145506} (\bibinfo
  {year} {2004})}\BibitemShut {NoStop}%
\bibitem [{\citenamefont {Lee}\ \emph {et~al.}(2007)\citenamefont {Lee},
  \citenamefont {Lim}, \citenamefont {Kim},\ and\ \citenamefont
  {Jin}}]{Lee_2007}%
  \BibitemOpen
  \bibfield  {author} {\bibinfo {author} {\bibfnamefont {J.~H.}\ \bibnamefont
  {Lee}}, \bibinfo {author} {\bibfnamefont {T.~K.}\ \bibnamefont {Lim}},
  \bibinfo {author} {\bibfnamefont {W.~T.}\ \bibnamefont {Kim}}, \ and\
  \bibinfo {author} {\bibfnamefont {J.~I.}\ \bibnamefont {Jin}},\ }\href@noop
  {} {\bibfield  {journal} {\bibinfo  {journal} {J Appl. Phys.}\ }\textbf
  {\bibinfo {volume} {101}},\ \bibinfo {pages} {034105} (\bibinfo {year}
  {2007})}\BibitemShut {NoStop}%
\bibitem [{\citenamefont {Liu}\ \emph {et~al.}(2016)\citenamefont {Liu},
  \citenamefont {Ackerman}, \citenamefont {Lubensky},\ and\ \citenamefont
  {Smalyukh}}]{Qliu_PNAS2016}%
  \BibitemOpen
  \bibfield  {author} {\bibinfo {author} {\bibfnamefont {Q.}~\bibnamefont
  {Liu}}, \bibinfo {author} {\bibfnamefont {P.~J.}\ \bibnamefont {Ackerman}},
  \bibinfo {author} {\bibfnamefont {T.~C.}\ \bibnamefont {Lubensky}}, \ and\
  \bibinfo {author} {\bibfnamefont {I.~I.}\ \bibnamefont {Smalyukh}},\
  }\href@noop {} {\bibfield  {journal} {\bibinfo  {journal} {PNAS}\ }\textbf
  {\bibinfo {volume} {113}},\ \bibinfo {pages} {10479} (\bibinfo {year}
  {2016})}\BibitemShut {NoStop}%
\bibitem [{\citenamefont {Brochard}\ and\ \citenamefont
  {de~Gennes}(1970)}]{Broc_1970}%
  \BibitemOpen
  \bibfield  {author} {\bibinfo {author} {\bibfnamefont {F.}~\bibnamefont
  {Brochard}}\ and\ \bibinfo {author} {\bibfnamefont {P.~G.}\ \bibnamefont
  {de~Gennes}},\ }\href@noop {} {\bibfield  {journal} {\bibinfo  {journal} {J.
  Phys.}\ }\textbf {\bibinfo {volume} {31}},\ \bibinfo {pages} {691} (\bibinfo
  {year} {1970})}\BibitemShut {NoStop}%
\bibitem [{\citenamefont {Mertelj}\ and\ \citenamefont
  {Lisjak}(2017)}]{Mert_LCR2017}%
  \BibitemOpen
  \bibfield  {author} {\bibinfo {author} {\bibfnamefont {A.}~\bibnamefont
  {Mertelj}}\ and\ \bibinfo {author} {\bibfnamefont {D.}~\bibnamefont
  {Lisjak}},\ }\href@noop {} {\bibfield  {journal} {\bibinfo  {journal} {Liq.
  Cryst. Rev.}\ }\textbf {\bibinfo {volume} {5}},\ \bibinfo {pages} {1}
  (\bibinfo {year} {2017})}\BibitemShut {NoStop}%
\bibitem [{\citenamefont {Mertelj}\ \emph {et~al.}(2013)\citenamefont
  {Mertelj}, \citenamefont {Lisjak}, \citenamefont {Drofenik},\ and\
  \citenamefont {{\v{C}}opi{\v{c}}}}]{Mert_Na2013}%
  \BibitemOpen
  \bibfield  {author} {\bibinfo {author} {\bibfnamefont {A.}~\bibnamefont
  {Mertelj}}, \bibinfo {author} {\bibfnamefont {D.}~\bibnamefont {Lisjak}},
  \bibinfo {author} {\bibfnamefont {M.}~\bibnamefont {Drofenik}}, \ and\
  \bibinfo {author} {\bibfnamefont {M.}~\bibnamefont {{\v{C}}opi{\v{c}}}},\
  }\href@noop {} {\bibfield  {journal} {\bibinfo  {journal} {Nature}\ }\textbf
  {\bibinfo {volume} {504}},\ \bibinfo {pages} {237} (\bibinfo {year}
  {2013})}\BibitemShut {NoStop}%
\bibitem [{\citenamefont {Mottram}\ and\ \citenamefont
  {Newton}(2014)}]{Mtram_ArX_2014}%
  \BibitemOpen
  \bibfield  {author} {\bibinfo {author} {\bibfnamefont {N.~J.}\ \bibnamefont
  {Mottram}}\ and\ \bibinfo {author} {\bibfnamefont {J.~P.}\ \bibnamefont
  {Newton}},\ }\href@noop {} {\bibfield  {journal} {\bibinfo  {journal} {arXiv
  preprint arXiv:1409.3542}\ } (\bibinfo {year} {2014})}\BibitemShut {NoStop}%
\bibitem [{\citenamefont {Bhattacharjee}(2010)}]{Amit_2010}%
  \BibitemOpen
  \bibfield  {author} {\bibinfo {author} {\bibfnamefont {A.}~\bibnamefont
  {Bhattacharjee}},\ }\emph {\bibinfo {title} {Inhomogeneous phenomena in
  nematic liquid crystals}},\ \href@noop {} {Ph.D. thesis} (\bibinfo {year}
  {2010})\BibitemShut {NoStop}%
\bibitem [{\citenamefont {Kaiser}\ \emph {et~al.}(1992)\citenamefont {Kaiser},
  \citenamefont {Wiese},\ and\ \citenamefont {Hess}}]{Kaiser_1992}%
  \BibitemOpen
  \bibfield  {author} {\bibinfo {author} {\bibfnamefont {P.}~\bibnamefont
  {Kaiser}}, \bibinfo {author} {\bibfnamefont {W.}~\bibnamefont {Wiese}}, \
  and\ \bibinfo {author} {\bibfnamefont {S.}~\bibnamefont {Hess}},\ }\href@noop
  {} {\bibfield  {journal} {\bibinfo  {journal} {J. Non-Equilib. Thermodyn.}\
  }\textbf {\bibinfo {volume} {17}},\ \bibinfo {pages} {153} (\bibinfo {year}
  {1992})}\BibitemShut {NoStop}%
\bibitem [{\citenamefont {Kralj}\ \emph {et~al.}(2010)\citenamefont {Kralj},
  \citenamefont {Rosso},\ and\ \citenamefont {Virga}}]{kralj_2010}%
  \BibitemOpen
  \bibfield  {author} {\bibinfo {author} {\bibfnamefont {S.}~\bibnamefont
  {Kralj}}, \bibinfo {author} {\bibfnamefont {R.}~\bibnamefont {Rosso}}, \ and\
  \bibinfo {author} {\bibfnamefont {E.~G.}\ \bibnamefont {Virga}},\ }\href@noop
  {} {\bibfield  {journal} {\bibinfo  {journal} {Physical Review E}\ }\textbf
  {\bibinfo {volume} {81}},\ \bibinfo {pages} {021702} (\bibinfo {year}
  {2010})}\BibitemShut {NoStop}%
\bibitem [{\citenamefont {Prost}\ and\ \citenamefont
  {de~Gennes}(1995)}]{JP_DG_1995}%
  \BibitemOpen
  \bibfield  {author} {\bibinfo {author} {\bibfnamefont {J.}~\bibnamefont
  {Prost}}\ and\ \bibinfo {author} {\bibfnamefont {P.~G.}\ \bibnamefont
  {de~Gennes}},\ }\href@noop {} {\emph {\bibinfo {title} {The Physics of Liquid
  Crystals}}},\ Vol.~\bibinfo {volume} {83}\ (\bibinfo  {publisher} {Oxford
  university press},\ \bibinfo {year} {1995})\BibitemShut {NoStop}%
\bibitem [{\citenamefont {Pleiner}\ \emph {et~al.}(2001)\citenamefont
  {Pleiner}, \citenamefont {Jarkova}, \citenamefont {Muler},\ and\
  \citenamefont {Brand}}]{HPlein_2001}%
  \BibitemOpen
  \bibfield  {author} {\bibinfo {author} {\bibfnamefont {H.}~\bibnamefont
  {Pleiner}}, \bibinfo {author} {\bibfnamefont {E.}~\bibnamefont {Jarkova}},
  \bibinfo {author} {\bibfnamefont {H.~W.}\ \bibnamefont {Muler}}, \ and\
  \bibinfo {author} {\bibfnamefont {H.~R.}\ \bibnamefont {Brand}},\ }\href@noop
  {} {\bibfield  {journal} {\bibinfo  {journal} {Magnetohydrodynamics}\
  }\textbf {\bibinfo {volume} {37}},\ \bibinfo {pages} {146} (\bibinfo {year}
  {2001})}\BibitemShut {NoStop}%
\bibitem [{\citenamefont {Bisht}\ \emph {et~al.}(2019)\citenamefont {Bisht},
  \citenamefont {Banerjee}, \citenamefont {Milewski},\ and\ \citenamefont
  {Majumdar}}]{konark_2019}%
  \BibitemOpen
  \bibfield  {author} {\bibinfo {author} {\bibfnamefont {K.}~\bibnamefont
  {Bisht}}, \bibinfo {author} {\bibfnamefont {V.}~\bibnamefont {Banerjee}},
  \bibinfo {author} {\bibfnamefont {P.}~\bibnamefont {Milewski}}, \ and\
  \bibinfo {author} {\bibfnamefont {A.}~\bibnamefont {Majumdar}},\ }\href@noop
  {} {\bibfield  {journal} {\bibinfo  {journal} {Phys. Rev. E}\ }\textbf
  {\bibinfo {volume} {100}},\ \bibinfo {pages} {012703} (\bibinfo {year}
  {2019})}\BibitemShut {NoStop}%
\bibitem [{\citenamefont {Bisht}\ \emph {et~al.}(2020)\citenamefont {Bisht},
  \citenamefont {Wang}, \citenamefont {Banerjee},\ and\ \citenamefont
  {Majumdar}}]{Konark_2019_2}%
  \BibitemOpen
  \bibfield  {author} {\bibinfo {author} {\bibfnamefont {K.}~\bibnamefont
  {Bisht}}, \bibinfo {author} {\bibfnamefont {Y.}~\bibnamefont {Wang}},
  \bibinfo {author} {\bibfnamefont {V.}~\bibnamefont {Banerjee}}, \ and\
  \bibinfo {author} {\bibfnamefont {A.}~\bibnamefont {Majumdar}},\ }\href@noop
  {} {\bibfield  {journal} {\bibinfo  {journal} {Phys. Rev. E}\ }\textbf
  {\bibinfo {volume} {101}},\ \bibinfo {pages} {022706} (\bibinfo {year}
  {2020})}\BibitemShut {NoStop}%
\bibitem [{\citenamefont {Vats}\ \emph {et~al.}(2020)\citenamefont {Vats},
  \citenamefont {Banerjee},\ and\ \citenamefont {Puri}}]{V_2020}%
  \BibitemOpen
  \bibfield  {author} {\bibinfo {author} {\bibfnamefont {A.}~\bibnamefont
  {Vats}}, \bibinfo {author} {\bibfnamefont {V.}~\bibnamefont {Banerjee}}, \
  and\ \bibinfo {author} {\bibfnamefont {S.}~\bibnamefont {Puri}},\ }\href@noop
  {} {\bibfield  {journal} {\bibinfo  {journal} {Europhys. Lett.}\ }\textbf
  {\bibinfo {volume} {128}},\ \bibinfo {pages} {66001} (\bibinfo {year}
  {2020})}\BibitemShut {NoStop}%
\bibitem [{\citenamefont {Vats}\ \emph {et~al.}(2021)\citenamefont {Vats},
  \citenamefont {Banerjee},\ and\ \citenamefont {Puri}}]{V_2021}%
  \BibitemOpen
  \bibfield  {author} {\bibinfo {author} {\bibfnamefont {A.}~\bibnamefont
  {Vats}}, \bibinfo {author} {\bibfnamefont {V.}~\bibnamefont {Banerjee}}, \
  and\ \bibinfo {author} {\bibfnamefont {S.}~\bibnamefont {Puri}},\ }\href@noop
  {} {\bibfield  {journal} {\bibinfo  {journal} {Soft Matter}\ }\textbf
  {\bibinfo {volume} {17}},\ \bibinfo {pages} {2659} (\bibinfo {year}
  {2021})}\BibitemShut {NoStop}%
\bibitem [{\citenamefont {Ravnik}\ and\ \citenamefont
  {{\v{Z}}umer}(2009)}]{Ravnik_2009}%
  \BibitemOpen
  \bibfield  {author} {\bibinfo {author} {\bibfnamefont {M.}~\bibnamefont
  {Ravnik}}\ and\ \bibinfo {author} {\bibfnamefont {S.}~\bibnamefont
  {{\v{Z}}umer}},\ }\href@noop {} {\bibfield  {journal} {\bibinfo  {journal}
  {Liquid Crystals}\ }\textbf {\bibinfo {volume} {36}},\ \bibinfo {pages}
  {1201} (\bibinfo {year} {2009})}\BibitemShut {NoStop}%
\bibitem [{\citenamefont {Priestly}(2012)}]{Priestly_2012}%
  \BibitemOpen
  \bibfield  {author} {\bibinfo {author} {\bibfnamefont {E.}~\bibnamefont
  {Priestly}},\ }\href@noop {} {\emph {\bibinfo {title} {Introduction to Liquid
  Crystals}}}\ (\bibinfo  {publisher} {Springer Science},\ \bibinfo {year}
  {2012})\BibitemShut {NoStop}%
\bibitem [{\citenamefont {Puri}\ and\ \citenamefont
  {Wadhawan}(2009)}]{Puri_2009}%
  \BibitemOpen
  \bibfield  {author} {\bibinfo {author} {\bibfnamefont {S.}~\bibnamefont
  {Puri}}\ and\ \bibinfo {author} {\bibfnamefont {V.}~\bibnamefont
  {Wadhawan}},\ }\href@noop {} {\emph {\bibinfo {title} {Kinetics of Phase
  Transitions}}}\ (\bibinfo  {publisher} {CRC Press},\ \bibinfo {year} {2009})\
  pp.\ \bibinfo {pages} {8--68}\BibitemShut {NoStop}%
\bibitem [{\citenamefont {Bray}(2002)}]{Bray_2002}%
  \BibitemOpen
  \bibfield  {author} {\bibinfo {author} {\bibfnamefont {A.~J.}\ \bibnamefont
  {Bray}},\ }\href@noop {} {\bibfield  {journal} {\bibinfo  {journal} {Adv.
  Phys.}\ }\textbf {\bibinfo {volume} {51}},\ \bibinfo {pages} {481} (\bibinfo
  {year} {2002})}\BibitemShut {NoStop}%
\bibitem [{\citenamefont {Hohenberg}\ and\ \citenamefont
  {Krekhov}(2015)}]{Hberg_2015}%
  \BibitemOpen
  \bibfield  {author} {\bibinfo {author} {\bibfnamefont {P.~C.}\ \bibnamefont
  {Hohenberg}}\ and\ \bibinfo {author} {\bibfnamefont {A.~P.}\ \bibnamefont
  {Krekhov}},\ }\href@noop {} {\bibfield  {journal} {\bibinfo  {journal} {Phys.
  Rep.}\ }\textbf {\bibinfo {volume} {572}},\ \bibinfo {pages} {1} (\bibinfo
  {year} {2015})}\BibitemShut {NoStop}%
\bibitem [{\citenamefont {Li}\ \emph {et~al.}(2006)\citenamefont {Li},
  \citenamefont {Buchnev}, \citenamefont {Cheon}, \citenamefont {Glushchenko},
  \citenamefont {Reshetnyak}, \citenamefont {Reznikov}, \citenamefont
  {Sluckin},\ and\ \citenamefont {West}}]{Li_2006}%
  \BibitemOpen
  \bibfield  {author} {\bibinfo {author} {\bibfnamefont {F.}~\bibnamefont
  {Li}}, \bibinfo {author} {\bibfnamefont {O.}~\bibnamefont {Buchnev}},
  \bibinfo {author} {\bibfnamefont {C.~I.}\ \bibnamefont {Cheon}}, \bibinfo
  {author} {\bibfnamefont {A.}~\bibnamefont {Glushchenko}}, \bibinfo {author}
  {\bibfnamefont {V.}~\bibnamefont {Reshetnyak}}, \bibinfo {author}
  {\bibfnamefont {Y.}~\bibnamefont {Reznikov}}, \bibinfo {author}
  {\bibfnamefont {T.~J.}\ \bibnamefont {Sluckin}}, \ and\ \bibinfo {author}
  {\bibfnamefont {J.~L.}\ \bibnamefont {West}},\ }\href@noop {} {\bibfield
  {journal} {\bibinfo  {journal} {Phys. Rev. Lett.}\ }\textbf {\bibinfo
  {volume} {97}},\ \bibinfo {pages} {147801} (\bibinfo {year}
  {2006})}\BibitemShut {NoStop}%
\bibitem [{\citenamefont {Lopatina}\ and\ \citenamefont
  {Selinger}(2009)}]{Lena_2009}%
  \BibitemOpen
  \bibfield  {author} {\bibinfo {author} {\bibfnamefont {L.~M.}\ \bibnamefont
  {Lopatina}}\ and\ \bibinfo {author} {\bibfnamefont {J.~V.}\ \bibnamefont
  {Selinger}},\ }\href@noop {} {\bibfield  {journal} {\bibinfo  {journal}
  {Phys. Rev. Lett.}\ }\textbf {\bibinfo {volume} {102}},\ \bibinfo {pages}
  {197802} (\bibinfo {year} {2009})}\BibitemShut {NoStop}%
\bibitem [{\citenamefont {Lopatina}\ and\ \citenamefont
  {Selinger}(2011)}]{Lena_2011}%
  \BibitemOpen
  \bibfield  {author} {\bibinfo {author} {\bibfnamefont {L.~M.}\ \bibnamefont
  {Lopatina}}\ and\ \bibinfo {author} {\bibfnamefont {J.~V.}\ \bibnamefont
  {Selinger}},\ }\href@noop {} {\bibfield  {journal} {\bibinfo  {journal}
  {Phys. Rev. E}\ }\textbf {\bibinfo {volume} {84}},\ \bibinfo {pages} {041703}
  (\bibinfo {year} {2011})}\BibitemShut {NoStop}%
\bibitem [{\citenamefont {Gorkunov}\ and\ \citenamefont
  {Osipov}(2011)}]{Gorkunov_2011}%
  \BibitemOpen
  \bibfield  {author} {\bibinfo {author} {\bibfnamefont {M.~V.}\ \bibnamefont
  {Gorkunov}}\ and\ \bibinfo {author} {\bibfnamefont {M.~A.}\ \bibnamefont
  {Osipov}},\ }\href@noop {} {\bibfield  {journal} {\bibinfo  {journal} {Soft
  Matter}\ }\textbf {\bibinfo {volume} {7}},\ \bibinfo {pages} {4348} (\bibinfo
  {year} {2011})}\BibitemShut {NoStop}%
\bibitem [{\citenamefont {Susser}\ \emph {et~al.}(2021)\citenamefont {Susser},
  \citenamefont {Kralj},\ and\ \citenamefont {Rosenblatt}}]{Adam_2021}%
  \BibitemOpen
  \bibfield  {author} {\bibinfo {author} {\bibfnamefont {A.~L.}\ \bibnamefont
  {Susser}}, \bibinfo {author} {\bibfnamefont {S.}~\bibnamefont {Kralj}}, \
  and\ \bibinfo {author} {\bibfnamefont {C.}~\bibnamefont {Rosenblatt}},\
  }\href@noop {} {\bibfield  {journal} {\bibinfo  {journal} {Soft matter}\
  }\textbf {\bibinfo {volume} {17}},\ \bibinfo {pages} {9616} (\bibinfo {year}
  {2021})}\BibitemShut {NoStop}%
\bibitem [{\citenamefont {Majumdar}(2010)}]{Apala_2010}%
  \BibitemOpen
  \bibfield  {author} {\bibinfo {author} {\bibfnamefont {A.}~\bibnamefont
  {Majumdar}},\ }\href@noop {} {\bibfield  {journal} {\bibinfo  {journal} {Eur.
  J. Appl. Math.}\ }\textbf {\bibinfo {volume} {21}},\ \bibinfo {pages} {181}
  (\bibinfo {year} {2010})}\BibitemShut {NoStop}%
\bibitem [{\citenamefont {Forest}\ \emph {et~al.}(2000)\citenamefont {Forest},
  \citenamefont {Wang},\ and\ \citenamefont {Zhou}}]{Forest_2000}%
  \BibitemOpen
  \bibfield  {author} {\bibinfo {author} {\bibfnamefont {M.~G.}\ \bibnamefont
  {Forest}}, \bibinfo {author} {\bibfnamefont {Q.}~\bibnamefont {Wang}}, \ and\
  \bibinfo {author} {\bibfnamefont {H.}~\bibnamefont {Zhou}},\ }\href@noop {}
  {\bibfield  {journal} {\bibinfo  {journal} {Phys. Rev. E}\ }\textbf {\bibinfo
  {volume} {61}},\ \bibinfo {pages} {6655} (\bibinfo {year}
  {2000})}\BibitemShut {NoStop}%
\bibitem [{\citenamefont {Hohenberg}\ and\ \citenamefont
  {Halperin}(1977)}]{Hohenberg_1977}%
  \BibitemOpen
  \bibfield  {author} {\bibinfo {author} {\bibfnamefont {P.~C.}\ \bibnamefont
  {Hohenberg}}\ and\ \bibinfo {author} {\bibfnamefont {B.~I.}\ \bibnamefont
  {Halperin}},\ }\href@noop {} {\bibfield  {journal} {\bibinfo  {journal} {Rev.
  Mod. Phys.}\ }\textbf {\bibinfo {volume} {49}},\ \bibinfo {pages} {435}
  (\bibinfo {year} {1977})}\BibitemShut {NoStop}%
\bibitem [{\citenamefont {Kincaid}\ and\ \citenamefont
  {Cheney}(2009)}]{kin_Num2009}%
  \BibitemOpen
  \bibfield  {author} {\bibinfo {author} {\bibfnamefont {D.}~\bibnamefont
  {Kincaid}}\ and\ \bibinfo {author} {\bibfnamefont {E.~W.}\ \bibnamefont
  {Cheney}},\ }\href@noop {} {\emph {\bibinfo {title} {Numerical Analysis:
  Mathematics of Scientific Computing}}},\ Vol.~\bibinfo {volume} {2}\
  (\bibinfo  {publisher} {American Mathematical Soc.},\ \bibinfo {year}
  {2009})\BibitemShut {NoStop}%
\end{thebibliography}%

\end{document}